\documentclass[referee]{raa}
\usepackage{graphicx,times}
\usepackage{natbib}
\usepackage{amssymb,amsmath}
\usepackage{multirow}
\usepackage{hyperref}
\bibpunct{(}{)}{;}{a}{}{,}

\hypersetup{pdftitle = The title of my PDF, pdfauthor = My name, pdfsubject= The subject, pdfkeywords = keyword1 keyword2 keyword3}
\hypersetup{colorlinks = true, linkcolor = green, anchorcolor = red, citecolor = blue, filecolor = red, pagecolor = red, urlcolor = red}

\newcommand{\lines}{$\lambda\lambda\ 4960,5008\,$}
\newcommand{\linesNe}{$\lambda\lambda\ 3869,3968\,$}

\newcommand{\oxygen}{\mbox{[O {\sc iii}]}\ }
\newcommand{\neon}{\mbox{[Ne {\sc iii}]}\ }
\newcommand{\sulphur}{\mbox{[S {\sc ii}]}\ }
\newcommand{\linesS}{$\lambda\lambda\ 6717,6731$}

\begin{document}

   \title{Measuring the Time Variation of the Fine-structure Constant with Quasars Detected by LAMOST}

 \volnopage{Vol.0 (20xx) No.0, 000--000}      
   \setcounter{page}{1}

   \author{Jin-Nan Wei\inst{1,2,3} \and Rui-Jie Chen\inst{1,4} \and Jun-Jie Wei\inst{1,2} \and Mart\'{i}n L\'{o}pez-Corredoira\inst{5,6,7} \\
   \and Xue-Feng Wu\inst{1,2}
   }

   \institute{Purple Mountain Observatory, Chinese Academy of Sciences, Nanjing 210023, China; {\it jjwei@pmo.ac.cn, xfwu@pmo.ac.cn}\\
        \and School of Astronomy and Space Sciences, University of Science and Technology of China, Hefei 230026, China\\
        \and Shanghai Astronomy Museum (branch of Shanghai Science \& Technology Museum), Shanghai 201306, China\\
        \and Department of Physics, Imperial College London, London SW7 2AZ, UK \\
        \and Instituto de Astrof\'{i}sica de Canarias, E-38205 La Laguna, Tenerife, Spain\\
        \and PIFI-Visiting Scientist 2023 of China Academy of Sciences at Purple Mountain Observatory, Nanjing 210023 and National Astronomical Observatories, Beijing 100012, China\\
        \and Departamento de Astrof\'{i}sica, Universidad de La Laguna, E-38206 La Laguna, Tenerife, Spain\\
\vs \no
   {\small Received~~20xx month day; accepted~~20xx~~month day}
}

\abstract{The \oxygen \lines emission lines in the optical spectra of galaxies and quasars have
been widely used to investigate the possible variation of the fine-structure constant $\alpha$ over
cosmic time. In this work, we utilize the Large Sky Area Multi-object Fiber Spectroscopic Telescope
(LAMOST) quasar survey, for the first time, to measure the relative $\alpha$ variation $\Delta\alpha/\alpha$
in time through the \oxygen doublet method. From the LAMOST Data Release 9 quasar catalog, we refine
a sample of 209 quasar spectra with strong and narrow \oxygen emission lines over a redshift range of $0<z<0.8$.
Analysis on all of the 209 spectra obtains $\Delta\alpha/\alpha = (0.5 \pm 3.7) \times 10^{-4}$,
which suggests that there is no evidence of varying $\alpha$ on the explored cosmological timescales.
Assuming a linear variation, the mean rate of change in $\Delta\alpha/\alpha$ is limited to be
$(-3.4 \pm 2.4)\times 10^{-13}$ $\mathrm{yr^{-1}}$ in the last 7.0 Gyr. While our LAMOST-based constraint
on $\Delta\alpha/\alpha$ is not competitive with those of the Sloan Digital Sky Survey (SDSS) quasar
observations, our analysis serves to corroborate the results of SDSS with another independent survey.
}

   \authorrunning{Wei et al.}            
   \titlerunning{Fine-structure Constant with LAMOST}  
   \maketitle

%
\section{Introduction}           
\label{sect:intro}

Fundamental physical constants are assumed to be universal and constant under the current Standard Model
of particle physics. However, some modern theories beyond the Standard Model predict that the fundamental
constants of nature may vary across space in time \citep{2017RPPh...80l6902M}. In the last few decades,
many efforts have been devoted to search for possible variations of these constants, either through
laboratory experiments or astrophysical observations (see \citealt{RevModPhys.75.403,2011LRR....14....2U}
for a review).

One of the particularly interesting fundamental constants is the fine-structure constant
$\alpha$, defined by $\alpha \equiv e^2/(4\pi\epsilon_{0}\hbar c)$, where $e$, $\epsilon_{0}$, $\hbar$,
and $c$ are the electron charge, permittivity of free space, reduced Planck constant, and speed of light
in vacuum, respectively. It is a dimensionless physical constant that characterizes
the strength of the electromagnetic force between electrically charged particles.
Any possible variations in $\alpha$ could indicate that the laws of physics are
not the same everywhere in the Universe or have changed over time. This challenges the assumption
of the constancy of physical laws, which is a cornerstone of modern physics. Moreover, variations
in $\alpha$ could imply the existence of new physics beyond the Standard Model, such as extra dimensions
or varying scalar fields. This would have profound implications for our understanding of the Universe.
By studying potential variations in $\alpha$, physicists and astronomers can examine the robustness of
the fundamental laws of nature, test alternative cosmological models that account for varying $\alpha$
\citep{2015PhRvD..91j3501M,2015PhLB..743..377M}, and potentially uncover new aspects of the Universe's
underlying structure \citep{2004MNRAS.349..291M}.

The fundamental constant $\alpha$ also quantifies the separation in the fine structure of atomic spectral lines
(e.g., \citealt{1999PhRvA..59..230D,RevModPhys.75.403}). Any relative $\alpha$ variation (i.e., $\Delta\alpha/\alpha$)
over time can therefore be measured directly by comparing the wavelengths of fine-structure splitting
of atomic lines at two different epochs. Astrophysical spectra invloving long look-back times have been
widely used to investigate the possible variation of $\alpha$. The first measurements on the $\alpha$
variation from astronomical spectroscopy reached an accuracy of $\Delta\alpha/\alpha \approx10^{-2}-10^{-3}$
\citep{1956Natur.178..688S,1965ApJ...142.1677B,1967PhRvL..19.1294B,1967ApJ...149L..11B}. Since then,
the methodologies for analyzing spectra and our understanding of systematic errors have improved significantly.
At present, there are two main methods to measure the relative separation of absorption lines in the spectra
of quasars, i.e., the alkali-doublet (AD) method \citep{1967ApJ...149L..11B} and the many-multiplet (MM) method
\citep{1999PhRvL..82..888D,PhysRevLett.82.884}.

In the AD method, the adopted quasar absorption lines were
mainly fine-structure doublet lines, such as C~{\sc iv}, N~{\sc v}, Mg~{\sc ii}, and Si~{\sc iv} (e.g.,
\citealt{1994A&AS..104...89P,1995ApJ...453..596C,2001MNRAS.327.1237M,2005A&A...430...47C}). The current best
constraints obtained using the AD method are those based on the analysis of Si~{\sc iv} absorption lines,
yielding $\Delta\alpha/\alpha=(1.5\pm4.3)\times 10^{-6}$ over a redshift range of $1.59\leq z \leq 2.92$
\citep{2005A&A...430...47C}. The MM method simultaneously analyzes all (or most) doublets of many atomic species,
thereby achieving a higher precision compared to the AD method. Using the MM method, some early works claimed
to have found tentative evidence for variation of $\alpha$ at a level of $\Delta\alpha/\alpha \sim (1-10)\times 10^{-6}$
(e.g., \citealt{2001MNRAS.327.1208M,2003MNRAS.345..609M,2001PhRvL..87i1301W,2011PhRvL.107s1101W}). However,
subsequent works indicated that such variations were likely to be caused by wavelength distortions and other
systematic effects (e.g, \citealt{2014MNRAS.445..128E,2015MNRAS.447..446W}). Recently, \cite{2022Sci...378..634M}
applied the MM method to the absorption spectra of nearby star twins within 50 pc, and found no variations
in $\alpha$ with a precision of $5.0\times10^{-8}$. Although more precise, the MM method still suffers from
a number of uncertainties (see \citealt{2022Univ....8..266W} for a recent review), which may arise from the
techniques for correcting wavelength distortion \citep{2017MNRAS.468.1568D}, the assumptions underlying
the Voigt fitting technique \citep{2004LNP...648..151L}, the technical details of profile fitting
\citep{2017MNRAS.468.1639B,2023MNRAS.521..850L}, the unconsidered systematic errors \citep{2021MNRAS.507...27L},
and other. These uncertainties may induce biases on the values of $\Delta\alpha/\alpha$ and underestimations
of their errors. Thus, different methods are encouraged to cross-check the results of the MM method.

In this work, we employ the method based on the \oxygen emission lines, first proposed by \cite{1965ApJ...142.1677B},
to constrain possible variations in $\alpha$. Since the \oxygen doublet method relies only on a pair of lines,
the limits on $\Delta\alpha/\alpha$ are not as stringent as those obtained with the MM method, but has the advantage
of being more transparent and less subject to systematics. The \oxygen \lines doublet lines originate in the
downward transitions from the same upper energy level of the same ion, so no assumptions on ionization state,
chemical composition, or distribution of energy levels are required in practice. The \oxygen doublet method
therefore represents an excellent alternative for measuring the $\alpha$ variation on a firm basis.
The $\Delta\alpha/\alpha$ constraints obtained by recent works based on the \oxygen emission lines are summarized
as follows. By analyzing 42 quasar spectra from the Early Data Release of Sloan Digital Sky Survey (SDSS),
\cite{2004ApJ...600..520B} derived $\Delta\alpha/\alpha =(0.7\pm1.4)\times10^{-4}$ over the range $0.16<z<0.8$.
\cite{2010ApJ...713...46G} obtained $\Delta\alpha/\alpha =(2.4\pm2.5)\times10^{-5}$ using 1568 quasar spectra
at $0.0<z<0.8$ from SDSS Data Release 6 (DR6). \cite{2014MNRAS.439L..70R} derived $\Delta\alpha/\alpha =(-2.1\pm1.6)\times10^{-5}$
using 2347 quasar spectra at $0.02<z<0.74$ from SDSS DR7. \cite{2015MNRAS.452.4153A} obtained
$\Delta\alpha/\alpha =(0.9\pm1.8)\times10^{-5}$ using 13,175 quasar spectra at $0.04<z<1.0$ from SDSS DR12.
\cite{2024MNRAS.527.4913L} analyzed 40 spectra of Ly $\alpha$ emitting galaxies and 46 spectra of quasars
at $1.09<z<3.73$ using the VLT/X-Shooter spectra publicly available, from which they yielded
$\Delta\alpha/\alpha =(-3\pm6)\times10^{-5}$. \cite{2024arXiv240403123J} measured $\Delta\alpha/\alpha =(2\sim3)\times10^{-5}$
by utilizing $\sim 110,000$ \oxygen emission-line galaxies at $0<z<0.95$ from the Dark Energy Spectroscopic Instrument.
\cite{2024arXiv240508977J} obtained $\Delta\alpha/\alpha =(0.4\pm0.7)\times10^{-4}$ using 572 JWST spectra
from 522 \oxygen emission-line galaxies at $3<z<10$.
In addition to the \oxygen emission lines, other emission doublets, such as \neon \linesNe and \sulphur \linesS,
have also been used to explore the $\alpha$ variation (e.g., \citealt{2010ApJ...713...46G,2015MNRAS.452.4153A}).
However, the limits of their accuracy are worse, because all these doublets in quasars are fainter than \oxygen
and some of them are affected by systematic errors.

Recently, the Large Sky Area Multi-object Fiber Spectroscopic Telescope (LAMOST) released the results of its 9 yr
quasar survey \citep{2023ApJS..265...25J}. Here, we use the latest LAMOST DR9 quasar sample, for the first time,
to measure the time variation of $\alpha$ through the \oxygen doublet method. Our analysis can serve to corroborate
previous results of SDSS with another independent survey, thereby discarding possible systematic errors in the
wavelength calibration of quasar spectra in SDSS. The rest of this paper is organized as follows.
In Section~\ref{sec:sample}, we introduce our quasar sample and spectroscopic data. Our resulting constraints on
$\Delta\alpha/\alpha$ are then presented in Section~\ref{sec:result}. Finally, a brief summary and discussions
are drawn in Section~\ref{sec:conclusion}.

\section{LAMOST Data and Wavelength Measurements}
\label{sec:sample}
In this section, we will first clarify why the \oxygen doublet can provide an ideal testbed for measuring the $\alpha$
variation. We will then describe the LAMOST quasar survey and the refined sample used for our analysis. Finally, we will
introduce the measurements of emission-line wavelengths in detail.

\subsection{\oxygen Doublet as a Testbed for Varying $\alpha$}
The variation in the fine-structure constant $\alpha$ can be measured through the wavelength separation of absorption
or emission multiplets in the quasar spectra as \citep{RevModPhys.75.403}
\begin{equation}
    \frac{\Delta\alpha}{\alpha}\left(z\right) = \frac{1}{2}\left\{\frac{\left[(\lambda_{2} - \lambda_{1})/(\lambda_{2} + \lambda_{1})\right]_{z}}{\left[(\lambda_{2} - \lambda_{1})/(\lambda_{2} + \lambda_{1})\right]_{0}} - 1\right\}\;,
\label{eq:Delta}
\end{equation}
where $\lambda_{1}$ and $\lambda_{2}$ are the shorter and longer wavelengths of the pairs of the doublet, and
the subscripts $0$ and $z$ stand for the wavelength values at redshift zero (laboratory values) and at redshift $z$,
respectively.

The present-day vacuum wavelengths of the \oxygen doublet lines are $\lambda_{1}(0)=4960.295$\,\AA\, and
$\lambda_{2}(0)=5008.240$\,\AA, respectively.\footnote{\url{http://physics.nist.gov/PhysRefData/ASD/lines_form.html}}
Concerning emission lines, the \oxygen doublet is the most suitable pair of lines for measuring $\Delta\alpha/\alpha$.
The reasons are as follows. First, the doublet lines have a wide wavelength separation,
$\Delta \lambda_{0}=[\lambda_{2} - \lambda_{1}]_{0}=47.945$\,\AA, representing almost one order of magnitude wider
than most of fine-structure doublets. Note that the sensitivity of $\Delta\alpha/\alpha$ is positively related to
the wavelength separation. For illustrative purposes, Equation~(\ref{eq:Delta}) can be approximated as
$\Delta\alpha/\alpha \approx 0.5 \times \epsilon/\Delta \lambda_{0}$, where
$\epsilon=\Delta \lambda_{z}/(1+z)-\Delta \lambda_{0}$ denotes the difference between the measured wavelength
separation at redshift $z$ in rest frame and the local one. It is obvious from this formula that a difference of
$\epsilon=0.01$\,\AA\, for \oxygen implies $\Delta\alpha/\alpha \approx 10^{-4}$. That is, a statistical or
systematic uncertainty of $0.01$\,\AA\, places a measuring precision of $\Delta\alpha/\alpha \approx 10^{-4}$.
Second, the \oxygen doublet often appears in quasar spectra with relatively high signal-to-noise ratio (S/N).
Compared to other doublets, it is easier to extract the wavelength values of the \oxygen lines, which is
crucial for the $\Delta\alpha/\alpha$ constraint.

\subsection{LAMOST Quasar Survey and Sample Selection}
\label{subsec:sample}
LAMOST, also called the Guoshoujing Telescope, is a special quasi-meridian reflecting Schmidt telescope located at
Xinglong Observatory, China \citep{1996ApOpt..35.5155W,2004ChJAA...4....1S,2012RAA....12.1197C,2012RAA....12..723Z}.
The available large focal surface is circular with a diameter of 1.75 m ($\sim5\degr$ field of view), 4000 fibers
are almost uniformly distributed over it. Each spectrum obtained by LAMOST is divided into two channels (blue and red)
whose wavelength coverage is 3700--5900\,\AA\, and 5700--9000\,\AA, respectively, with an overlapping region at
5700--5900\,\AA. The spectra have a resolution of $R\sim$1000--2000 over the entire wavelength range.

After the two year commissioning period, a pilot spectroscopic survey with LAMOST was conducted between 2011 October
and 2012 June \citep{2012RAA....12.1243L}. The LAMOST regular survey officially begins on 2012 September, which
consists of two main tasks \citep{2012RAA....12..723Z}: the LAMOST Experiment for Galactic Understanding and
Exploration survey (LEGUE), and the LAMOST ExtraGAlactic Survey (LEGAS). The LAMOST quasar survey is affiliated with
LEGAS. Despite only a small portion of the observation time was used to search for quasars due to the limitations
of the observation site (e.g., bad weather, poor seeing, and bright background), a total of 56,175 quasars have
already been identified by LAMOST, 24,127 of which were newly discovered, during the first 9 yr quasar survey
\citep{2023ApJS..265...25J}.

In this work, we make use of the LAMOST low resolution catalog of emission line features of quasars to investigate
the possible variation of $\alpha$ over cosmic time. All the quasar spectra used for our analysis are downloaded
from LAMOST's official website.\footnote{\url{https://www.lamost.org/dr9/v2.0/catalogue}} In order to effectively
refine the final sample from the LAMOST DR9 catalog, our sample selection criteria include the following aspects.

(i) Those quasars with redshifts $\leq 0.8$ are selected. This restriction is imposed by the wavelength range of
the LAMOST spectrograph (3700--9000\,\AA) and the wavelength positions of the \oxygen $\lambda\lambda4960,\,5008$
doublet lines. This criterion removes the sample down to 16,902 quasars.

(ii) Those targets with strong \oxygen emission lines are selected. Since the $\lambda4960$ line is always weaker
than the $\lambda5008$ line, the selection of the final sample is determined mainly on the basis of the strength
of the $\lambda4960$ line. We require the peak flux density of the weaker \oxygen line (4960\,\AA) to be larger
than $10^{-19}\,\mathrm{erg}\,\mathrm{s^{-1}\,\mathrm{cm^{-2}}}$ \AA$^{-1}$\,and its $\mathrm{S/N}_{\oxygen\,4960}$
to be above 3. This criterion significantly reduces the sample from 16,902 to 373 objects.

(iii) Those targets with high goodness of fit are selected. Owing to the low resolution mode of the LAMOST
quasar survey, there are some scenarios in which the Gaussian fits to the spectral lines does not converge.
This criterion causes us to further discard 164 spectra. Hence, there are 209 remaining quasar spectra
in our final sample.

\subsection{Measurements of Emission-line Wavelengths}
The emission-line properties of the \oxygen doublet can be measured by fitting the LAMOST released spectra.
Following \cite{2023ApJS..265...25J}, we adopt the publicly available Python code (\emph{\sc PyQSOFit};
\citealt{2018ascl.soft09008G}) and its extended package (\emph{\sc QSOFITMORE}; \citealt{yuming_fu_2021_5810042})
to fit the spectra. With the estimated uncertainties of the pixels that derived from the reduction pipeline,
the \emph{\sc PyQSOFit} code performs the $\chi^2$ fits. Before the fitting, each quasar spectrum should be corrected
for Galactic extinction using the reddening map of \cite{1998ApJ...500..525S} with an extinction curve of
$R_{V}=3.1$ \citep{1999PASP..111...63F}. After the extinction correction, the spectrum is then
transformed into the rest frame by using the redshift $z$.

We estimate the continuum by fitting a broken power law ($f_{\rm bpl}$) and an iron model ($f_{\mathrm{Fe\,{\sc II}}}$)
to the rest-frame spectrum, masking those wavelength windows that contain quasar emission lines and
the LAMOST spectral overlapping region. Based on the analysis result of the mean composite quasar spectra
obtained by \cite{2001AJ....122..549V}, we fix the inflection point of the broken power law at $4661$ \AA\;in rest frame.
Many previous studies have found that a sudden slope change occurs at $\sim5000$ \AA\; in the quasar continuum
(e.g., \citealt{1985ApJ...288...94W,2001AJ....122..549V}). There are two possible reasons for the steeper slope
at longer wavelengths. One probable reason is the near-infrared inflection caused by hot dust emission
\citep{1994ApJS...95....1E}. Another reason may be the contamination from low-redshift host galaxies,
which would contribute a larger proportion at longer wavelengths \citep{1998MNRAS.301....1S,2001AJ....122..549V}.
Besides the broken power law model, the Fe\,{\sc ii} model $f_{\mathrm{Fe\,{\sc II}}}$ is also an important
component of the continuum template, i.e.,
\begin{equation}
    f_{\mathrm{Fe\,{\sc II}}} = b_{0}F_{\mathrm{Fe\,{\sc II}}}(\lambda,\,b_{1},\,b_{2})\;,
\end{equation}
where $b_{0}$ is the normalization, $b_{1}$ represents the full width at half maximum (FWHM) of Gaussian profile
applied to convolve the Fe\,{\sc ii} template, and $b_{2}$ denotes the wavelength shift acted on the
Fe\,{\sc ii} template. A detailed description of the Fe\,{\sc ii} template can be found in \cite{2023ApJS..265...25J}.
Most of the quasar continuum can be well described by the broken power law plus an Fe\,{\sc ii} template,
but some spectra have strange shapes in their continuum. This problem may arise from some uncertainties
in the spectral response curve, which are occasionally caused by poor relative flux calibrations and
unstable efficiencies of some fibers. To overcome this problem, we add an additional three-order polynomial
model ($f_{\mathrm{poly}}$; \citealt{2020ApJS..249...17R,2022ApJS..261...32F}).
It is thus clear that the pseudocontinuum would be fitted by two (or three) components:
\begin{equation}
    f_{\mathrm{cont}} = f_{\rm bpl} + f_{\mathrm{Fe\,{\sc II}}} + (f_{\mathrm{poly}})\;.
\end{equation}

\begin{figure*}
\centering
\begin{tabular}{c}
{(a)}\\
\includegraphics[width=12cm]{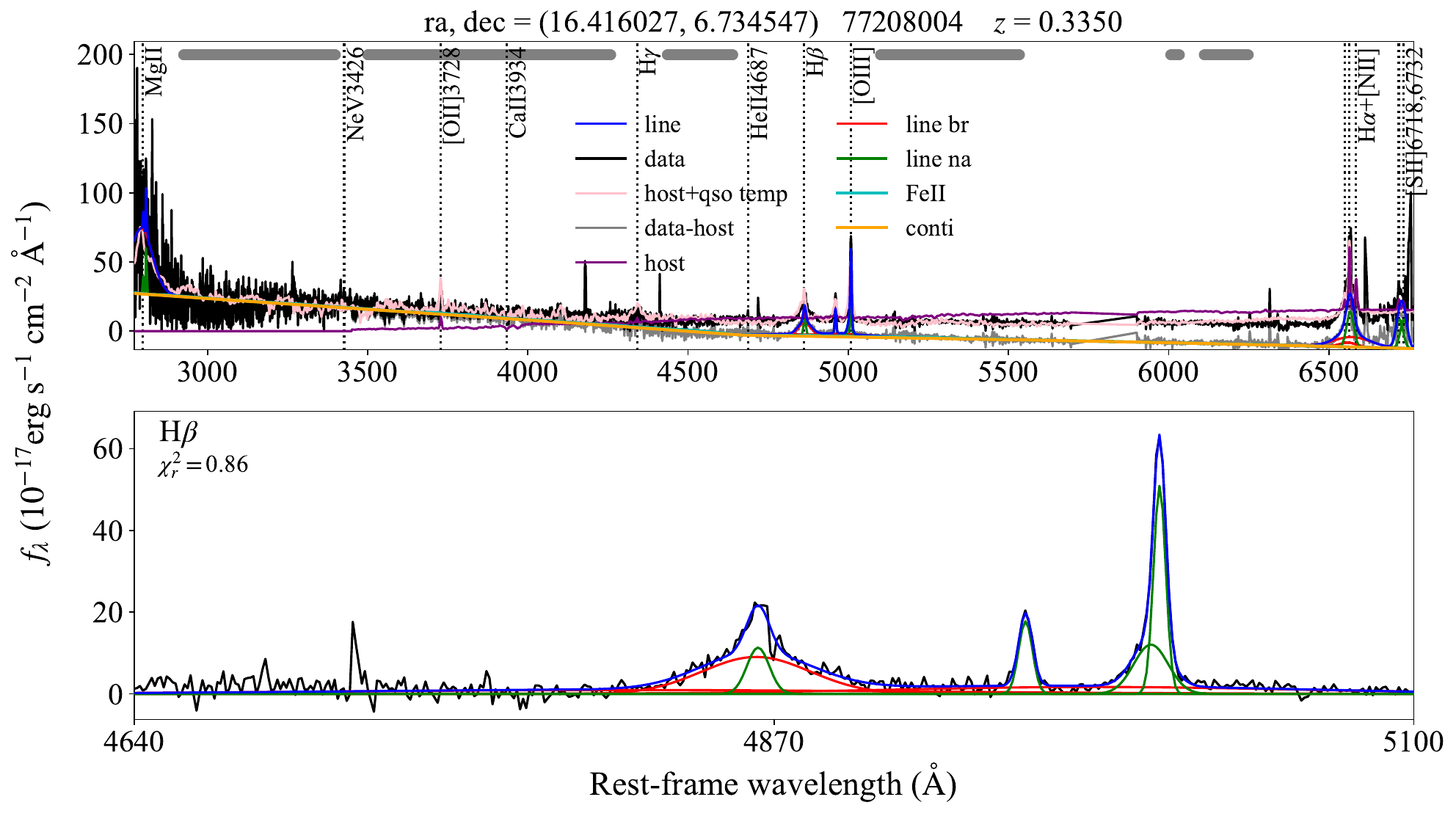} \\
{(b)}\\
\includegraphics[width=12cm]{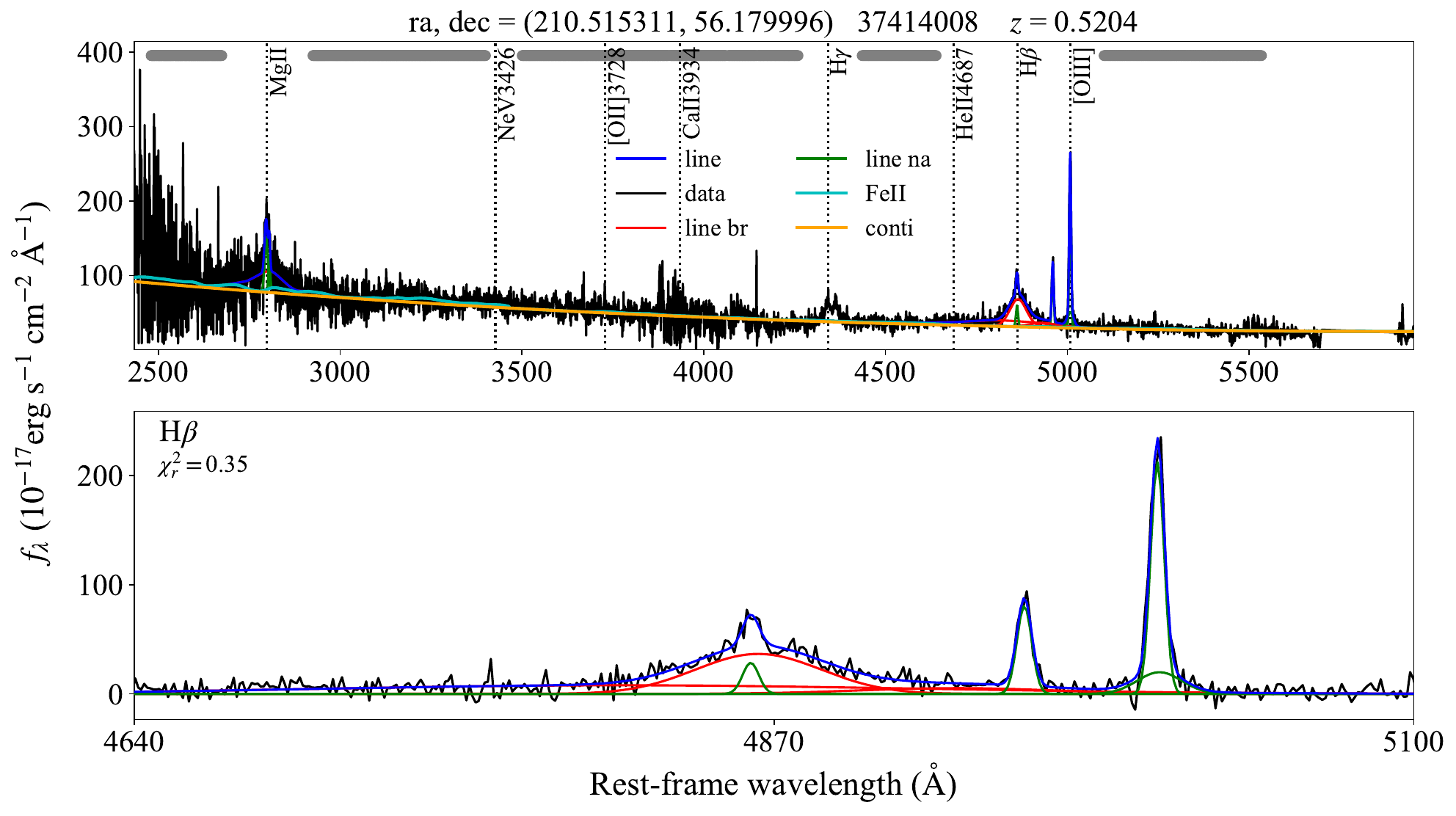}
\end{tabular}
\caption{Two examples of the spectral fitting results with (panel (a)) and without (panel (b))
the host galaxy template for quasars whose ID numbers are 77208004 and 37414008. In each pair
of panels, the upper panel shows the fitting results of the whole spectrum: the black lines represent
the dereddened spectra, the yellow lines denote the continuum model of $(f_{\rm bpl}+f_{\mathrm{poly}})$,
the cyan lines denote the Fe\,{\sc ii} template, the purple line represents the host galaxy component,
and the gray line represents the dereddened spectrum with the decomposition of the host galaxy.
In each pair of panels, the lower panel shows the deblending results of the H$\beta$--\oxygen emission lines:
the black lines represent the extinction-corrected spectra with the continuum and the host galaxy contamination
(if it exists) subtracted. As for the fitted emission lines, the broad and narrow components are marked in red and green, respectively, and their sums are in blue. The goodness-of-fit $\chi_{r}^{2}$ are also listed.}
\label{fig1}
\end{figure*}

It is worth emphasizing that one should check whether the quasar spectral data is contaminated by the host galaxy,
before fitting the continuum. In general, for high-$z$ ($z\ga0.5$) or high-luminosity ($\log_{10}L_{5100}\ga44.5$)
quasars, the contamination from the host galaxy is negligible. While for those low-$z$ or low-luminosity quasars,
the host galaxy contributes an average of $\sim15\%$ of the observed emissions and produces a $\sim0.06$ dex
overestimate of the continuum luminosity at 5100\,\AA\;\citep{2011ApJS..194...45S}. The spectral fitting packages
\emph{\sc PyQSOFit} and \emph{\sc QSOFITMORE} would automatically determine whether the quasar spectrum is
contaminated by the host galaxy.
If true, the decomposition of the host galaxy would be applied to the spectra. The decomposition is based on the
host galaxy template developed by \cite{2004AJ....128..585Y,2004AJ....128.2603Y}. Examples of the fitting results
with and without the host galaxy template are presented in Figure~\ref{fig1}(a) and (b), respectively. The coordinate,
observation ID, and redshift of each quasar are shown on the top of each plot. It is clear that for the quasar
with ID number 77208004, the host contribution can lead to overestimate of the strength of the emission lines
at longer wavelengths (see the top panel of Figure~\ref{fig1}(a)).

After subtracting the fitted continuum component and the host galaxy contamination (if it exists) from the spectrum,
the leftover emission-line components can be fitted with Gaussian profiles. The emission lines of H$\beta$ (narrow
and broad components) and \oxygen $\lambda\lambda4960,\,5008$ within the rest-frame window [4640, 5100]\,\AA\,are
simultaneously fitted. The broad component of H$\beta$ is modeled by two Gaussian profiles, and its narrow component
is fitted by a single Gaussian. Here we are used to setting the upper limit of FWHM for the narrow components to be
900 $\mathrm{km\,s^{-1}}$, which is the FWHM criterion for distinguishing between the narrow and broad components
\citep{2009ApJ...707.1334W,2019A&A...625A.123C,2019ApJ...882....4W}. In principle, the contribution of the broad
H$\beta$ emission line could produce a blueshift in the estimation of the \oxygen line positions, particularly affecting
the weak 4960\,\AA\,line. That is, in addition to a narrow component, the \oxygen $\lambda\lambda4960,\,5008$
doublet lines should include blue wing components (e.g., \citealt{2004A&A...413.1087C,2018A&A...615A..13S}). Therefore,
both of the \oxygen $\lambda\lambda4960,\,5008$ lines are modeled by two Gaussian profiles, one for the line core
and the other for the blueshifted wing, and neither of them is correlated with the narrow component of H$\beta$.
The wavelength position of each line of the \oxygen doublet can then be directly estimated as the central position
of the corresponding Gaussian for the line core. Examples of the best-fitting results of the H$\beta$--\oxygen
emission lines are given in the bottom panels of Figure~\ref{fig1}(a) and (b). The resulting goodness-of-fit
values of reduced $\chi_{r}^{2}$ are also shown in the figures.

\section{Results on the $\alpha$ variation}
\label{sec:result}
\begin{figure}
    \centering
    \includegraphics[width=8.5cm]{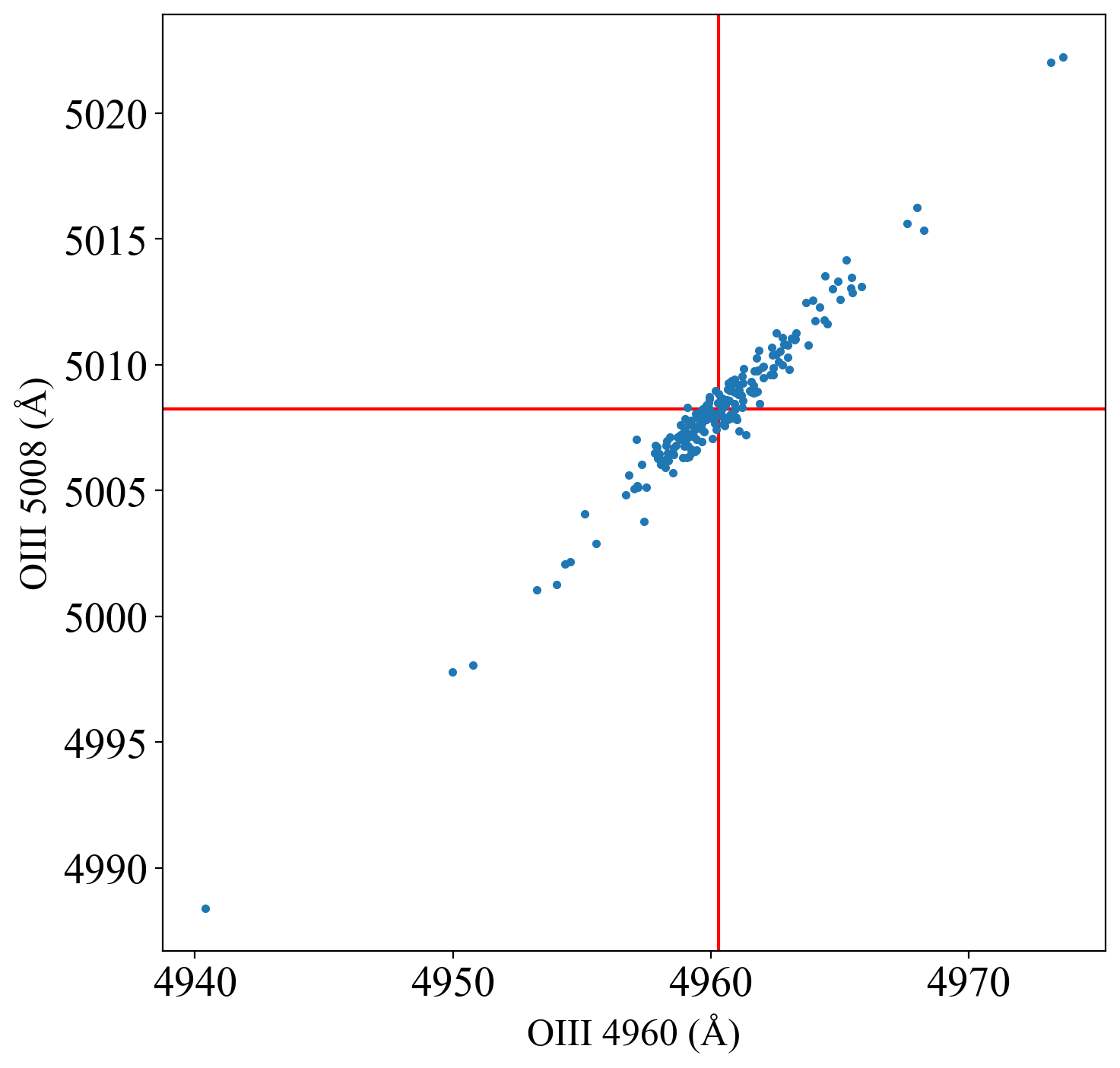}
    \caption{Rest-frame wavelength positions of the two emission lines of the \oxygen doublet in our final sample.
    The vertical and horizontal lines stand for the theoretical local values.}
    \label{fig:wavelength-position}
\end{figure}

We use a total of 209 quasar spectra, drawn from the LAMOST DR9 catalog, after applying the selection criteria
(i)-(iii) (see Section~\ref{subsec:sample}), to measure $\Delta\alpha/\alpha$. With the wavelength measurements
of the \oxygen \lines doublet lines for each quasar spectra, we calculate $\Delta\alpha/\alpha$ using
Equation~(\ref{eq:Delta}). Our results show that most of the $\Delta\alpha/\alpha$ measurements are consistent with
$0$ within $3\sigma$ confidence level, and the accuracies of $\Delta\alpha/\alpha$ are between
$10^{-4}$ and $10^{-2}$. Figure~\ref{fig:wavelength-position} shows the rest-frame wavelength measurements
of the \oxygen doublet lines of all 209 quasar spectra in our final sample. One can see from this plot that
the wavelengths of the two lines are aligned along a line from bottom left to top right, showing no systematic effects.

With a series of measured values $x_{i}=(\Delta\alpha/\alpha)_{i}$, we calculate the weighted average for
the final sample through
\begin{equation}
\overline{x}=\frac{\sum_{i}x_{i}/\sigma^{2}_{i}}{\sum_{i}1/\sigma^{2}_{i}}\;,
\label{eq:average}
\end{equation}
where $\sigma_{i}$ is the error of $x_{i}$. The corresponding uncertainty on $\overline{x}$ can be obtained from
\begin{equation}
\sigma^{2}_{\overline{x}}=\frac{1}{\sum_{i}1/\sigma^{2}_{i}}\;.
\label{eq:error}
\end{equation}
The weighted average for all the 209 spectra is $\Delta\alpha/\alpha=(0.5 \pm 3.7)\times 10^{-4}$. This value
is compatible with previous results obtained using other observational samples with the same method
\citep{2004ApJ...600..520B,2024arXiv240508977J}.

\begin{figure}
    \centering
    \includegraphics[width=10cm]{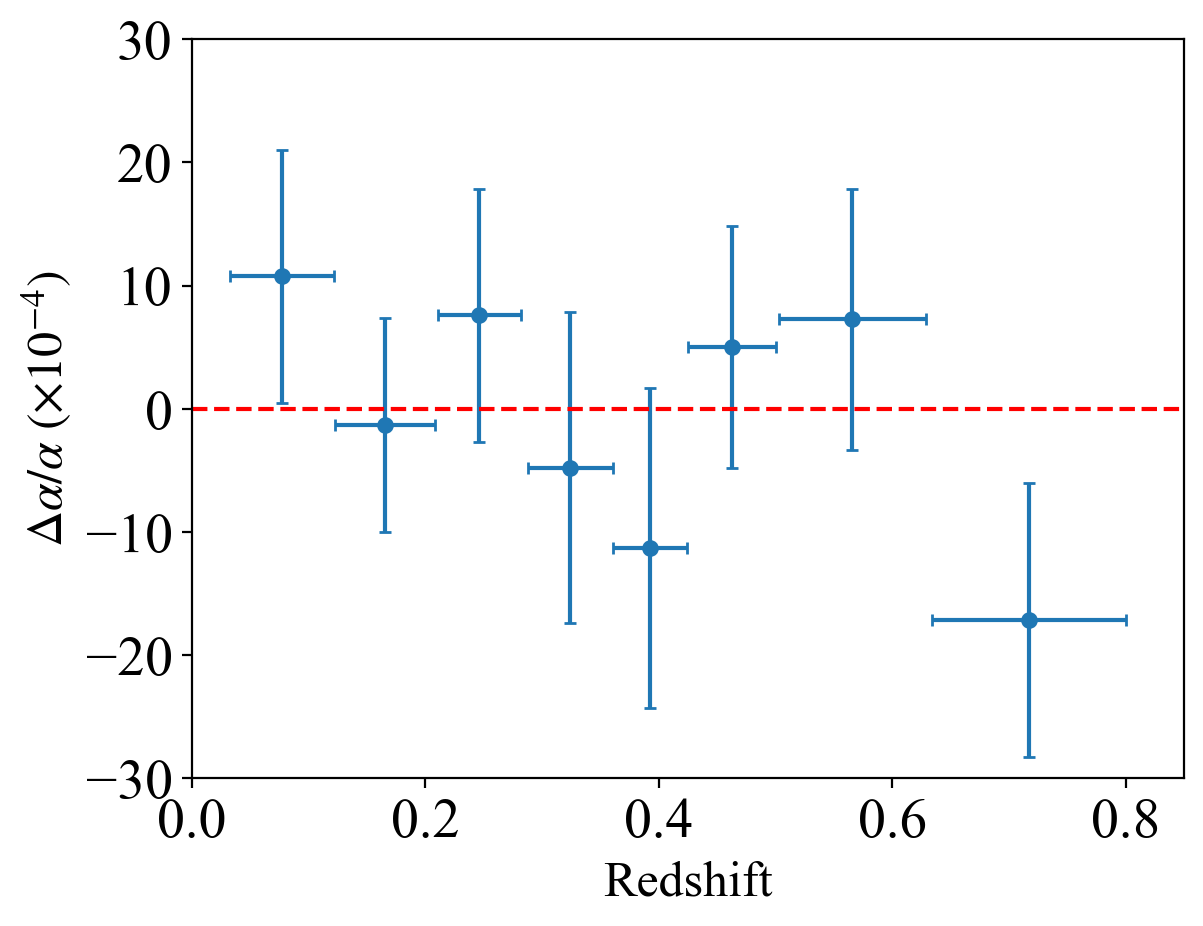}
    \caption{The redshift dependence of the $\Delta\alpha/\alpha$ measurements. Each redshift bin contains
    the contribution of $\sim 26$ quasar spectra.}
    \label{fig:timevariation}
\end{figure}

To explore the possible time variation of $\Delta\alpha/\alpha$, we divide the final sample into eight subsamples
with redshifts from low to high, with each subsample containing approximately the same number ($N\simeq26$) of spectra.
For each subsample, we also compute the average value of $\Delta\alpha/\alpha$ and its uncertainty through
Equations~(\ref{eq:average}) and (\ref{eq:error}). The averages of $\Delta\alpha/\alpha$ as a function of redshift
(or look-back time) are shown in Figure~\ref{fig:timevariation} and Table~\ref{tab1}. We do not find any variation
of $\Delta\alpha/\alpha$ over cosmic time, because none of the $\Delta\alpha/\alpha$ averages deviate from $0$ by
more than $1.5\sigma$ confidence level

The redshift range of quasars in our final sample is between 0.033 and 0.8, corresponding to a look-back time
($t_{\rm LB}$) of 0.5--7.0 Gyr, or the age of the Universe of 6.8--13.3 Gyr. We assume that $\alpha$ shows
linear change with time:
\begin{equation}
\Delta \alpha/\alpha=\kappa\cdot t_{\rm LB}+\omega\;,
\end{equation}
where $\kappa$ and $\omega$ are two free parameters.
A linear fit of $\Delta\alpha/\alpha$ with respect to $t_{\rm LB}$ for all 209 spectra gives a slope of
$\kappa=(-3.4 \pm 2.4)\times 10^{-13}$ $\mathrm{yr^{-1}}$ and an intercept of $\omega=(1.3 \pm 1.0)\times 10^{-3}$.
Here the slope $\kappa=\mathrm{d}(\Delta\alpha/\alpha)/\mathrm{d}t_{\rm LB}$ refers to the mean rate of change in
$\Delta\alpha/\alpha$ \citep{2004ApJ...600..520B}.

\begin{table}
\centering \caption{Average $\Delta\alpha/\alpha$ for different redshift intervals}
\begin{tabular}{ccc}
\hline
\hline
   Redshift interval      &   Number   &  $\Delta\alpha/\alpha\,(\times10^{-4})$    \\
\hline
0.033--0.122  &  26  &  $10.8 \pm 10.3$  \\
0.123--0.208  &  26  &  $-1.3 \pm 8.7$  \\
0.211--0.282  &  26  &  $7.6 \pm 10.3$  \\
0.288--0.361  &  26  &  $-4.8 \pm 12.7$  \\
0.361--0.425  &  26  &  $-11.3 \pm 13.0$  \\
0.425--0.500  &  26  &  $5.0 \pm 9.8$  \\
0.503--0.629  &  26  &  $7.3 \pm 10.6$  \\
0.634--0.801  &  27  &  $-17.1 \pm 11.1$  \\
\hline
\end{tabular}
\label{tab1}
\end{table}

We emphasize that all our results are based on the \oxygen emission lines. In principle,
by comparing the multiple absorption lines in the damped Ly$\alpha$ systems of quasar spectra, the MM
method can achieve higher accuracy than the \oxygen doublet method. As the MM method simultaneously
analyzes all absorption lines, it requires accurate measurements of both the line wavelengths in the
observed spectra and the laboratory wavelengths for all involved atomic transitions. Due to the low
resolution of LAMOST, the absorption lines of different atoms are difficult to extract accurately
from the observed quasar spectra. Therefore, it is hard to apply the MM method to the quasar sample
used here.

\section{Summary and Discussions}
\label{sec:conclusion}

In this work, we have used the LAMOST quasar survey, for the first time, to constrain the possible time variation of
the fine-structure constant $\alpha$ through the \oxygen doublet method. A great advantage of \oxygen is that its
doublet lines have a wide wavelength separation, which makes it very sensitive to the measurement of the relative
$\alpha$ variation $\Delta\alpha/\alpha$. The other advantage is that the \oxygen emission lines are much stronger
than any other doublets in many quasars, which is crucial for the $\Delta\alpha/\alpha$ measurement.

From 56,175 objects identified as quasars in the LAMOST DR9 quasar catalog, we have extracted a sample of 209 quasars
with strong \oxygen emission lines up to redshift 0.8. With this refined sample, we estimated a weighted average value
of $\Delta\alpha/\alpha = (0.5 \pm 3.7) \times 10^{-4}$ during the last 7.0 Gyr. Due to the smaller number of quasars
and the lower resolution of LAMOST with respect to SDSS, our measuring precision of $\Delta\alpha/\alpha$ is worse than
previous results obtained using different SDSS quasar samples with the same method by one order of magnitude
\citep{2010ApJ...713...46G,2014MNRAS.439L..70R,2015MNRAS.452.4153A}. While our LAMOST-based constraint is not competitive,
there is merit to the result. Our analysis serves to confirm the results of SDSS with another independent survey,
so we can exclude possible systematic errors in the wavelength calibration of spectra in SDSS.

To analyze the value of $\Delta\alpha/\alpha$ as a function of redshift, we divided the sample into eight redshift
bins, with each bin containing approximately the same number of quasars. We found that the averages of $\Delta\alpha/\alpha$
in all redshift bins are consistent with 0 within $1.5\sigma$ confidence level. This indicates that there is no evidence
of changes in $\Delta\alpha/\alpha$ with redshift. We limited the mean rate of change in $\Delta\alpha/\alpha$ to be
$(-3.4 \pm 2.4)\times 10^{-13}$ $\mathrm{yr^{-1}}$ within the last 7.0 Gyr.

To achieve better constraints on $\Delta\alpha/\alpha$ ($<10^{-6}$) using the emission-line method, high-resolution
spectroscopy ($R\sim 100,000$) is required. The measurement of $\Delta\alpha/\alpha$ using the LAMOST low resolution
spectra is doomed to be unable to reach the best precision from previous quasar observations. Nonetheless,
the LAMOST ongoing survey is still collecting useful data, and much more valuable quasars are expected
to be identified in the future. The LAMOST quasar survey not only provides an independent measurement of
$\Delta\alpha/\alpha$, but also helps to cross-check the results of other surveys. This work only
focused on the \oxygen doublet, but we will consider more emission lines (e.g., \sulphur) in future research.
In addition, we plan to discuss collaborative efforts with other observatories to combine datasets,
thereby enhancing statistical power for constraining variations in $\alpha$. For example, there are
a total of 56,175 identified quasars in the LAMOST DR9 quasar catalog, of which 24,127 are newly
discovered and not reported by SDSS \citep{2023ApJS..265...25J}. We plan to combine the SDSS quasar
survey with those new ones discovered by LAMOST to achieve a more robust constraint on $\Delta\alpha/\alpha$.


\normalem
\begin{acknowledgements}
We are grateful to the anonymous referee for the  helpful comments.
We thank Ye Li for her kind help on the download of the LAMOST data.
This work is partially supported by the National Natural Science Foundation of China
(grant Nos. 12422307, 12373053, and 12321003), the Key Research Program of Frontier Sciences
(grant No. ZDBS-LY-7014) of Chinese Academy of Sciences, and the Natural Science
Foundation of Jiangsu Province (grant No. BK20221562). M.L.C. is supported by
Chinese Academy of Sciences President's International Fellowship Initiative
(grant No. 2023VMB0001).
\end{acknowledgements}

\bibliographystyle{raa}
\bibliography{bibtex}




\end{document}